%% file: main.tex
\documentclass[conference,a4paper]{IEEEtran}
\usepackage{xcolor}
\usepackage{balance}
\usepackage{cite}
\usepackage{multirow}
\usepackage[pdftex]{graphicx}
\usepackage{amsmath,amssymb,amsfonts}
\usepackage{textcomp}
\usepackage[caption=false,font=footnotesize]{subfig}
\usepackage[nolist]{acronym}
\usepackage[capitalize]{cleveref}
\usepackage{nicefrac}
\usepackage{graphicx}  % remove 'demo' option for your real document
\crefname{table}{Tab.}{Tabs.}
\Crefname{table}{Tab.}{Tabs.}
\crefformat{equation}{(#2#1#3)}
\newcommand\norm[1]{\left\lVert#1\right\rVert}

\begin{document}
\input{./acronyms.tex}
\title{Deep Learning-based mmWave MIMO Channel Estimation using sub-6\,GHz Channel Information: CNN and UNet Approaches}

\author{\IEEEauthorblockN{
Faruk Pasic\IEEEauthorrefmark{1}\thanks{This work has been funded by the Christian Doppler Laboratory for Digital Twin assisted AI for sustainable Radio Access Networks, Institute of Telecommunications, TU Wien. The financial support by the Austrian Federal Ministry for Labour and Economy and the National Foundation for Research, Technology and Development and the Christian Doppler Research Association is gratefully acknowledged.},
Lukas Eller\IEEEauthorrefmark{1}\IEEEauthorrefmark{2},
Stefan Schwarz\IEEEauthorrefmark{1},
Markus Rupp\IEEEauthorrefmark{1} and
Christoph F. Mecklenbräuker\IEEEauthorrefmark{1}
}%

\IEEEauthorblockA{\IEEEauthorrefmark{1}% 2nd affiliations
Institute of Telecommunications, TU Wien, Vienna, Austria}
\IEEEauthorblockA{\IEEEauthorrefmark{2}% 2nd affiliations
Christian Doppler Laboratory for Digital Twin assisted AI for sustainable Radio Access Networks}
\IEEEauthorblockA{faruk.pasic@tuwien.ac.at}
}

\IEEEoverridecommandlockouts 

\makeatletter
\def\thanks#1{\protected@xdef\@thanks{\@thanks
        \protect\footnotetext{#1}}}
\makeatother

\maketitle

\begin{abstract}
Future wireless \ac{MIMO} systems will integrate both sub-6\,GHz and \ac{MMW} frequency bands to meet the growing demands for high data rates.
\ac{MIMO} link establishment typically requires accurate channel estimation, which is particularly challenging at \ac{MMW} frequencies due to the low \ac{SNR}.
In this paper, we propose two novel deep learning-based methods for estimating \ac{MMW} \ac{MIMO} channels by leveraging out-of-band information from the sub-6\,GHz band.
The first method employs a \ac{CNN}, while the second method utilizes a UNet architecture.
We compare these proposed methods against deep-learning methods that rely solely on in-band information and with other state-of-the-art out-of-band aided methods.
Simulation results show that our proposed out-of-band aided deep-learning methods outperform existing alternatives in terms of achievable spectral efficiency.
\end{abstract}
\vskip0.5\baselineskip

\acresetall

\section{Introduction}
\Ac{MMW} communications have the potential to meet high data rate demands due to their extensive bandwidth~\cite{Ai2020, Pasic2023_mag}.
To overcome high propagation losses and ensure sufficient link margin, most \ac{MMW} systems will employ large antenna arrays and \ac{MIMO} techniques~\cite{Heath2016}.
The key challenge in implementing \ac{MMW} \ac{MIMO} systems is link establishment, which relies on accurate channel estimation~\cite{Heath2016}.
However, at \ac{MMW} frequencies, channel estimation is particularly challenging due to the low pre-beamforming \ac{SNR}.

Recently, \ac{ML} techniques based on the multi-layer perceptrons, commonly referred as deep learning, have attracted significant attention for \ac{MMW} \ac{MIMO} channel estimation~\cite{Dong2019, Jin2019, Zhao2023}.
Authors have utilized deep \ac{CNN}~\cite{Dong2019}, fast and flexible denoising \ac{CNN} (FFDNet)~\cite{Jin2019}, and ResNet-UNet structure~\cite{Zhao2023} to address channel estimation in \ac{MMW} \ac{MIMO} systems.
However, these methods rely solely on in-band \ac{MMW} channel information, which is inherently constrained by low pre-beamforming \ac{SNR}, thereby limiting their overall performance.

Furthermore, \ac{MMW} systems are being implemented in conjunction with sub-6\,GHz systems to facilitate multi-band communication~\cite{Shafi2020}.
Multiple multi-band measurement campaigns in various environments have revealed that sub-6\,GHz and \ac{MMW} frequency bands share similar multipath characteristics~\cite{Pasic2022, Hofer2024, Pasic2023, Radovic2023}.
However, systems operating below 6\,GHz exhibit lower propagation loss, resulting in a higher pre-beamforming \ac{SNR}.
This reliable out-of-band information from sub-6\,GHz can be utilized to support and enhance channel estimation for \ac{MMW} systems.
The use of out-of-band information for \ac{MMW} channel estimation has been investigated in~\cite{Pasic2024, Pasic2024_TCOM}.
These studies propose novel pilot-aided channel estimation methods for \ac{MMW} \ac{MIMO} systems, utilizing the \ac{LOS} channel component acquired with the support of the sub-6\,GHz band.
However, without incorporating \ac{ML}, many inherent relationships between in-band and out-of-band channel estimates remain unexplored.

\textbf{Contribution:}
In this paper, we improve the out-of-band aided channel estimation methods from~\cite{Pasic2024_TCOM} by incorporating \ac{ML}.
Unlike conventional methods, \ac{ML} has a greater capacity to uncover the intrinsic relationships between in-band and out-of-band channel estimates.
We propose two novel methods for \ac{MMW} \ac{MIMO} systems: one based on a \ac{CNN} and the other employing a UNet architecture.
We evaluate the proposed methods through simulations in terms of normalized channel \ac{MSE} and \ac{SE}.

\textbf{Organization:}
\cref{sec:system_model} presents the system model.
\cref{sec:methods} outlines the channel estimation methods, followed by their simulation-based evaluation in~\cref{sec:comparison}.
Finally,~\cref{sec:conclusion} concludes the paper.

\textbf{Notation:} 
The superscript $\left( \cdot \right) ^{\left( \rm b \right)}$ represents frequency-band dependent values, where ${\rm b} \in \{ {\rm s}, {\rm m} \}$, with ${\rm s}$ referring to the sub-6\,GHz and ${\rm m}$ to the \ac{MMW} band.
Scalars are denoted by $x$, vectors by bold lowercase letters ${\mathbf x}$ and matrices by bold uppercase letters ${\mathbf X}$.
The $i$-th row of the matrix ${\mathbf X}$ is indicated by $\mathbf{X}_{i, :}$, and the $j$-th column by $\mathbf{X}_{:, j}$.
The all-ones matrix is denoted by $\mathbf{1}$, with its dimensions specified in the subscript.
The superscripts $\left( \cdot \right) ^{\rm T}$ and $\left( \cdot \right) ^{\rm H}$ represent transpose and Hermitian transpose, respectively.
The Euclidean norm is represented by $\lVert \cdot \rVert$ and the Frobenius norm by $\lVert \cdot \rVert_F$.

\section{System Model} \label{sec:system_model}
\begin{figure}
    \centering
    \includegraphics[width=\linewidth]{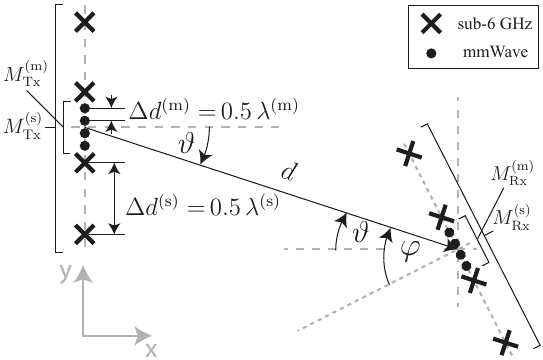}
    \caption{A point-to-point MIMO system with co-located sub-6\,GHz and \ac{MMW} antenna arrays. 
    The system geometry is characterized by the number of transmit antennas $M_{\rm Tx}^{\left( \rm b \right)}$, receive antennas $M_{\rm Rx}^{\left( \rm b \right)}$ and their mutual separation $\Delta d^{\left( \rm b \right)}$, along with the corresponding \ac{AoD} $\vartheta$ and \ac{AoA} $\varphi$.}
    \label{fig:system_model}
\end{figure}

We consider a point-to-point \ac{MIMO} system that operates simultaneously on both sub-6\,GHz and \ac{MMW} frequency bands in the radiative far-field regime.
Both the sub-6\,GHz and \ac{MMW} arrays are arranged in a \ac{ULA} configuration, co-located and aligned to ensure identical \ac{AoD} $\vartheta$ and \ac{AoA} $\varphi$ for both arrays (see~\cref{fig:system_model}).
We assume perfect synchronization between the transmitters and receivers in both time and frequency domains.
Each transmitter and receiver is equipped with a single \ac{RF} chain per antenna, enabling fully digital beamforming at both the sub-6\,GHz and \ac{MMW} frequency bands.
The system operates using a \ac{TDD} protocol, ensuring reciprocal channel responses.
The transmission system employs an \ac{OFDM} system and quadrature amplitude modulation with $N^{\left( \rm b \right)}$ subcarriers.

\subsection{Channel Model} \label{subsec:channel_model}
At the $n$-th \ac{OFDM} subcarrier, the channel is represented by the matrix $\mathbf{H}^{\left( \rm b \right)} [n] \in \mathbb{C}^{M_{\rm Rx}^{\left( \rm b \right)}  \times M_{\rm Tx}^{\left( \rm b \right)} }$, based on the equivalent \ac{OFDM} complex baseband representation.
The channel follows a frequency-selective Rician fading channel model~\cite{molisch2012wireless}
\begin{equation}
    \begin{split}    
        \mathbf{H}^{\left( \rm b \right)} [n] = & 
         \sqrt{\eta^{\left( \rm b \right)}} 
         \sqrt{\frac{K^{\left( \rm b \right)}}{1+K^{\left( \rm b \right)}}} 
         \underbrace{\mathbf{H}_{\rm fs}^{\left( \rm b \right)}}_{\rm free-space} \\ 
         + & \sqrt{\eta^{\left( \rm b \right)}}
        \sqrt{\frac{1}{ 1+K^{\left( \rm b \right)}} } 
        \underbrace{\mathbf{H}_{\rm rp}^{\left( \rm b \right)} [n]}_{\rm Rayleigh\, part},
    \end{split}
    \label{eq:channel_model}
\end{equation}
where $\eta^{\left( \rm b \right)}$ represents the path gain including shadowing and $K^{\left( \rm b \right)}$ indicates the Rician $K$-factor for the band ${\rm b}$.
Given the different multipath characteristics between \ac{MMW} and sub-6\,GHz bands, we assume that the $K$-factors for these bands are related by $K^{\left( \rm m \right)}= c_K K^{\left( \rm s \right)}$, where $c_K$ is the scaling factor.

The deterministic \ac{LOS} channel $\mathbf{H}_{\rm fs}^{\left( \rm b \right)} [n]$ is given by
\begin{equation}
    \mathbf{H}_{\rm fs}^{\left( \rm b \right)} = 
    e^{j \chi^{\rm \left( b \right)}}
    \mathbf{a}_{\rm Rx}^{\rm \left( b \right)} \left( \varphi  \right) 
    \left( \mathbf{a}_{\rm Tx}^{\rm \left( b \right)} \left( \vartheta  \right) \right) ^{\rm H},
    \label{eq:free_space_component}
\end{equation}
where
\begin{equation}
    \mathbf{a}_{\rm Tx}^{\rm \left( b \right)} \left( \vartheta \right) =   
    \begin{bmatrix}
        1 & \cdots  & e^{-j 2 \pi \left( M_{\rm Tx}^{\rm \left( b \right)} - 1 \right) \frac{\Delta d^{\left( \rm b \right)}}{\lambda^{\rm \left( b \right)}} \sin \left( \vartheta \right) } 
        \end{bmatrix}^{\rm T}
\end{equation}
and
\begin{equation}
    \mathbf{a}_{\rm Rx}^{\rm \left( b \right)} \left( \varphi \right) =     
    \begin{bmatrix}
        1 & \cdots  & e^{-j 2 \pi \left( M_{\rm Rx}^{\rm \left( b \right)} - 1 \right) \frac{\Delta d^{\rm \left( b \right)}}{\lambda^{\rm \left( b \right)}} \sin \left( \varphi \right) }     
        \end{bmatrix}^{\rm T}
\end{equation}
are the steering vectors corresponding to the \ac{AoD} $\vartheta$ and the \ac{AoA} $\varphi$, respectively.
In~\cref{eq:free_space_component}, $\chi^{\rm \left( b \right)} ~\sim~\mathcal{U}~\left( - \pi, \pi \right)$ denotes the initial random phase offset of the free-space \ac{LOS} channel.

The Rayleigh matrix $\mathbf{H}_{\rm rp}^{\left( \rm b \right)} [n]$ is modeled following the \ac{3GPP} \ac{MIMO} channel model~\cite{3gpp.38.901}.
Due to the significant differences between sub-6\,GHz and \ac{MMW} wavelengths, the diffuse components $\mathbf{H}_{\rm rp}^{\left( \rm s \right)} [n]$ and $\mathbf{H}_{\rm rp}^{\left( \rm m \right)} [n]$ are assumed to be completely uncorrelated.
The channel matrix $\mathbf{H}^{\left( \rm b \right)}[\tau]$ in the delay domain for the delay-tap $\tau$ is then derived by applying the \ac{IDFT} to $\mathbf{H}^{\left( \rm b \right)} [n]$.

\subsection{Link Establishment} \label{subsec:link_establishment}
Link establishment involves a training phase followed by data transmission. 
In the training phase, we first estimate the wireless channel at both the sub-6\,GHz and \ac{MMW} frequency bands. 
These channel estimates are then combined and utilized for data transmission over the \ac{MMW} band.

\subsubsection{Training Phase} \label{subsubsec:training_phase}
The training phase consists of two sequential steps. 
In the first step, we transmit non-precoded pilot symbols across both frequency bands.
The input-output relationship for this training step is expressed as
\begin{equation} 
    \mathbf{y}^{\left( \rm b \right)} [n] = \mathbf{H}^{\left( \rm b \right)} [n] \boldsymbol{\phi}^{\left( \rm b \right)} [n] + \mathbf{w}^{\left( \rm b \right)} [n],
    \label{eq:training_input_output}
\end{equation}
where $\mathbf{y}^{\left( \rm b \right)} [n] \in \mathbb{C}^{M_{\rm Rx}^{\left( \rm b \right)} \times 1} $ denotes the received symbols and $\mathbf{w}^{\left( \rm b \right)} [n] \in \mathcal{CN}(0,\sigma_{w^{\left( \rm b \right)}}^2 \mathbf{I}_{M_{\rm Rx}^{\left( \rm b \right)}})$ represents the \ac{AWGN}.
In~\cref{eq:training_input_output}, the pilot symbols for the $t$-th transmit antenna are defined as follows:
\begin{equation}
    \mathrm{\phi}_t^{\left( \rm b \right)} [n] = 
    \begin{cases}
        \mathrm{\phi}^{\left( \rm b \right)} [n], \quad n \in \{ t, t+M_{\rm Tx}^{\left( \rm b \right)}, \ldots, N^{\left( \rm b \right)} \}
        \\
        0, \quad \quad \quad \text{else}
    \end{cases}
     \label{eq:pilot_allocation}
\end{equation}
At the receiver, the channel estimate $\widetilde{\mathbf{H}}^{\left( \rm b \right)} [n] \in \mathbb{C}^{M_{\rm Rx}^{\left( \rm b \right)} \times M_{\rm Tx}^{\left( \rm b \right)}}$ is obtained using \ac{LS} estimation, followed by linear interpolation.
To enable optimal beamforming matrix selection, the channel estimated at the receiver must also be available at the transmitter. 
Given that the system operates using the \ac{TDD} protocol, channel reciprocity in both frequency bands is utilized to enable acquisition of the channel estimate at the transmitter.

In the second step, the sub-6\,GHz channel estimate $\widetilde{\mathbf{H}}^{\left( \rm s \right)} [n]$ is used to jointly estimate the \ac{AoA} and \ac{AoD} of the \ac{LOS} component using the estimation method described in~\cite{Pasic2024_TCOM}.
Using this estimated angular information $\widetilde{\varphi}$ and $\widetilde{\vartheta}$, we transmit precoded/combined pilot symbols in the direction of the \ac{LOS} component within the \ac{MMW} band only.
The input-output relationship for this step is given by
\begin{equation} 
    \begin{split}
    \mathrm{y}^{\left( \rm m \right)} [n] & = 
    \underbrace{
    \left( \mathbf{a}_{\rm Rx}^{\rm \left( m \right)} \left( \widetilde{\varphi}  \right) \right) ^ {\rm H}
    \mathbf{H}^{\left( \rm m \right)} [n] \, 
    \mathbf{a}_{\rm Tx}^{\rm \left( m \right)} \left( \widetilde{\vartheta}  \right) }_
    {\mathrm{G}^{\left( \rm m \right)} [n]}
    \phi^{\left( \rm m \right)} [n] \\
    & + 
    \left( \mathbf{a}_{\rm Rx}^{\rm \left( m \right)} \left( \widetilde{\varphi}  \right) \right) ^ {\rm H}
    \mathbf{w}^{\left( \rm m \right)} [n],
    \end{split}
    \label{eq:training_input_output_2}
\end{equation}
where $\phi^{\left( \rm m \right)} [n]$ and $\mathrm{y}^{\left( \rm m \right)} [n]$ are one-dimensional complex-valued variables, given that only a single spatial stream is used.
Next, \ac{LS} estimation is applied, followed by linear interpolation, to obtain the effective beamformed scalar channel estimate $\widehat{\mathrm{G}}^{\left( \rm m \right)} [n]$. 
The \ac{LOS} component of $\widehat{\mathrm{G}}^{\left( \rm m \right)} [n]$ is then isolated through delay-domain filtering $ \widehat{\mathrm{G}}_{\rm LOS}^{\left( \rm m \right)} [\tau] = \widehat{\mathrm{G}}^{\left( \rm m \right)} [\tau]  \delta [\tau]$,
where $\delta [\cdot]$ denotes the delta function.
Finally, the \ac{LOS} channel estimate $\widehat{\mathrm{G}}_{\rm LOS}^{\left( \rm m \right)} [\tau]$ is transformed back to the frequency domain via the \ac{DFT}.
The resulting out-of-band aided channel estimate for the \ac{LOS} component is reconstructed as
\begin{equation}
    \widehat{\mathbf{H}}^{\left( \rm m \right)} [n] = 
     \mathbf{a}_{\rm Rx}^{\rm \left( m \right)} \left( \widetilde{\varphi}  \right)
    \widehat{\mathrm{G}}_{\rm LOS}^{\left( \rm m \right)} [n]
    \left( \mathbf{a}_{\rm Tx}^{\rm \left( m \right)} \left( \widetilde{\vartheta}  \right) \right) ^ {\rm H}.
    \label{eq:los_channel_estimation}
\end{equation}

\subsubsection{Data Transmission} \label{subsec:data_transmission}
We first process the channel estimates obtained during the training phase using the methods described in~\cref{sec:methods}.
The resulting \ac{MMW} channel estimate $\overline{\mathbf{H}}^{\left( \rm m \right)} [n]$ is then used for precoding and combining. 
To achieve optimal performance in terms of \ac{SE}, we apply \ac{SVD} to $\overline{\mathbf{H}}^{\left( \rm m \right)} [n]$, with its compact form given by
\begin{equation} 
    \overline{\mathbf{H}}^{\left( \rm m \right)} [n] = 
    \overline{\mathbf{Q}}^{\left( \rm m \right)} [n]  
    \overline{\mathbf{\Sigma}}^{\left( \rm m \right)} [n]
    \left( \overline{\mathbf{F}}^{\left( \rm m \right)} [n] 
    \right) ^ {\rm H}.
    \label{eq:svd}
\end{equation}
In~\cref{eq:svd}, $\overline{\mathbf{Q}}^{\left( \rm m \right)} [n] \in \mathbb{C}^{M_{\rm Rx}^{\left( \rm m \right)} \times {\ell_{\rm max}}}$ is the matrix of left singular vectors, $\overline{\mathbf{F}}^{\left( \rm m \right)} [n] \in \mathbb{C}^{M_{\rm Tx}^{\left( \rm m \right)} \times {\ell_{\rm max}}}$ is the matrix of right singular vectors, $\ell_{\rm max} = {\rm min} \left( M_{\rm Rx}^{\left( \rm m \right)}, M_{\rm Tx}^{\left( \rm m \right)} \right)$ is
the maximum number of streams and $\overline{\mathbf{\Sigma}}^{\left( \rm m \right)} [n]$ is the diagonal matrix of singular values $ \overline{\sigma}_{(1)}^{\left( \rm m \right)} [n], \ldots , \overline{\sigma}_{(\ell_{\rm max})}^{\left( \rm m \right)} [n]$.
The power loading matrix, characterized by diagonal elements $\overline{p}_{(1)}^{\left( \rm m \right)} [n], \ldots , \overline{p}_{(\ell_{\rm max})}^{\left( \rm m \right)} [n] $, is optimized to maximize the transmission rate by following the water-filling power allocation policy~\cite{Tse2005}.
Further, the precoding matrix ensures compliance with the total transmit power constraint
\begin{equation}
    \norm{ 
    \overline{\mathbf{F}}^{\left( \rm m \right)} [n] 
    \left( \overline{\mathbf{P}}^{\left( \rm m \right)} [n] \right)^{1/2} 
    }_F^2 = P_{\rm T}^{\left( \rm m \right)}.    
    \label{eq:power_normalization}
\end{equation}
The input-output relationship for the data transmission phase is given by
\begin{equation} 
    \begin{split}
    \mathbf{y}^{\left( \rm m \right)} [n] & =
    \left( \overline{\mathbf{Q}}^{\left( \rm m \right)} [n] \right) ^ {\rm H}
    \mathbf{H}^{\left( \rm m \right)} [n]
    \overline{\mathbf{F}}^{\left( \rm m \right)} [n]
    \left( \overline{\mathbf{P}}^{\left( \rm m \right)} [n] \right)^{1/2}
    \mathbf{x}^{\left( \rm m \right)} [n] \\
    & + \left( \overline{\mathbf{Q}}^{\left( \rm m \right)} [n] \right) ^ {\rm H}
    \mathbf{w}^{\left( \rm m \right)} [n],        
    \end{split}
    \label{eq:data_input_output}
\end{equation}
where $\mathbf{x}^{\left( \rm m \right)} [n] \in \mathbb{C}^{\ell_{\rm max} \times 1}$ denotes the transmit symbol vector, $\mathbf{y}^{\left( \rm m \right)} [n] \in \mathbb{C}^{\ell_{\rm max} \times 1}$ represents the received symbol vector and $\mathbf{w}^{\left( \rm m \right)} [n] \in \mathcal{CN}(0,\sigma_{w^{\left( \rm m \right)}}^2 \mathbf{I}_{M_{\rm Rx}^{\left( \rm m \right)}})$ is the \ac{AWGN}.

\section{Channel Estimation Methods} \label{sec:methods}
In this section, we introduce channel estimation methods for the \ac{MMW} system, combining in-band $\widetilde{\mathbf{H}}^{\left( \rm m \right)} [n]$ and out-of-band aided $\widehat{\mathbf{H}}^{\left( \rm m \right)} [n]$ channel estimates to improve estimation quality.
We first revisit the \ac{MRC} method, which is the best-performing method from~\cite{Pasic2024_TCOM} in terms of \ac{SE}.
Following this, we propose two novel channel estimation methods that combine the available estimates using \ac{ML} algorithms.

\subsection{Maximal Ratio Combining (MRC)} \label{subsec:mrc}
This method combines the out-of-band aided channel estimate $\widehat{\mathbf{H}}^{\left( \rm m \right)} [n]$ and the in-band estimate $\widetilde{\mathbf{H}}^{\left( \rm m \right)} [n]$ using \ac{MRC}, given by
\begin{equation}
    \overline{\mathbf{H}}^{\left( {\rm m} \right)} [n] = 
    \widehat{w} \,
    \widehat{\mathbf{H}}^{\left( {\rm m} \right)} [n]
    + (1-\widehat{w})
    \widetilde{\mathbf{H}}^{\left( {\rm m} \right)} [n],
   \label{eq:mrc}
\end{equation} 
where 
\begin{equation}
    \widehat{w} \approx
    \frac{ M_{\rm Tx}^{\left( \rm m \right)} M_{\rm Rx}^{\left( \rm m \right)} \sigma_{w^{\left( \rm m \right)}}^2}
    { \frac{M_{\rm Tx}^{\left( \rm m \right)} M_{\rm Rx}^{\left( \rm m \right)}}{1 + c_K \widetilde{K}^{\left( \rm s \right)}}
    + \left( 1 + M_{\rm Tx}^{\left( \rm m \right)} M_{\rm Rx}^{\left( \rm m \right)} \right) \sigma_{w^{\left( \rm m \right)}}^2 }.
    \label{eq:optimal_w}
\end{equation}
denotes the approximated optimal combining factor, with its detailed derivation provided in~\cite{Pasic2024_TCOM}.
In~\cref{eq:optimal_w}, $\widetilde{K}^{\left( \rm s \right)}$ represents the $K$-factor for the sub-6\,GHz band, estimated using the method of moments~\cite{Greenstein1999}.
This approach is designed to adapt to varying wireless channel conditions, particularly in relation to the $K$-factor and the noise variance $\sigma_{w^{\left( \rm m \right)}}^2$.

\subsection{Deep Learning-based Methods} \label{subsec:deep_learning}
With the aim of improving the performance of the proposed \ac{MRC} method in terms of \ac{SE}, we apply \ac{CNN}-based and UNet-based channel estimation, exploiting in-band and out-of-band aided channel estimates.
The deployment of both the \ac{CNN} and UNet consists of two phases: the offline training and the online deployment.

\subsubsection{CNN-based Method} \label{subsubsec:cnn_method}
\begin{figure}[t]
    \centering
    \includegraphics[width=\linewidth]{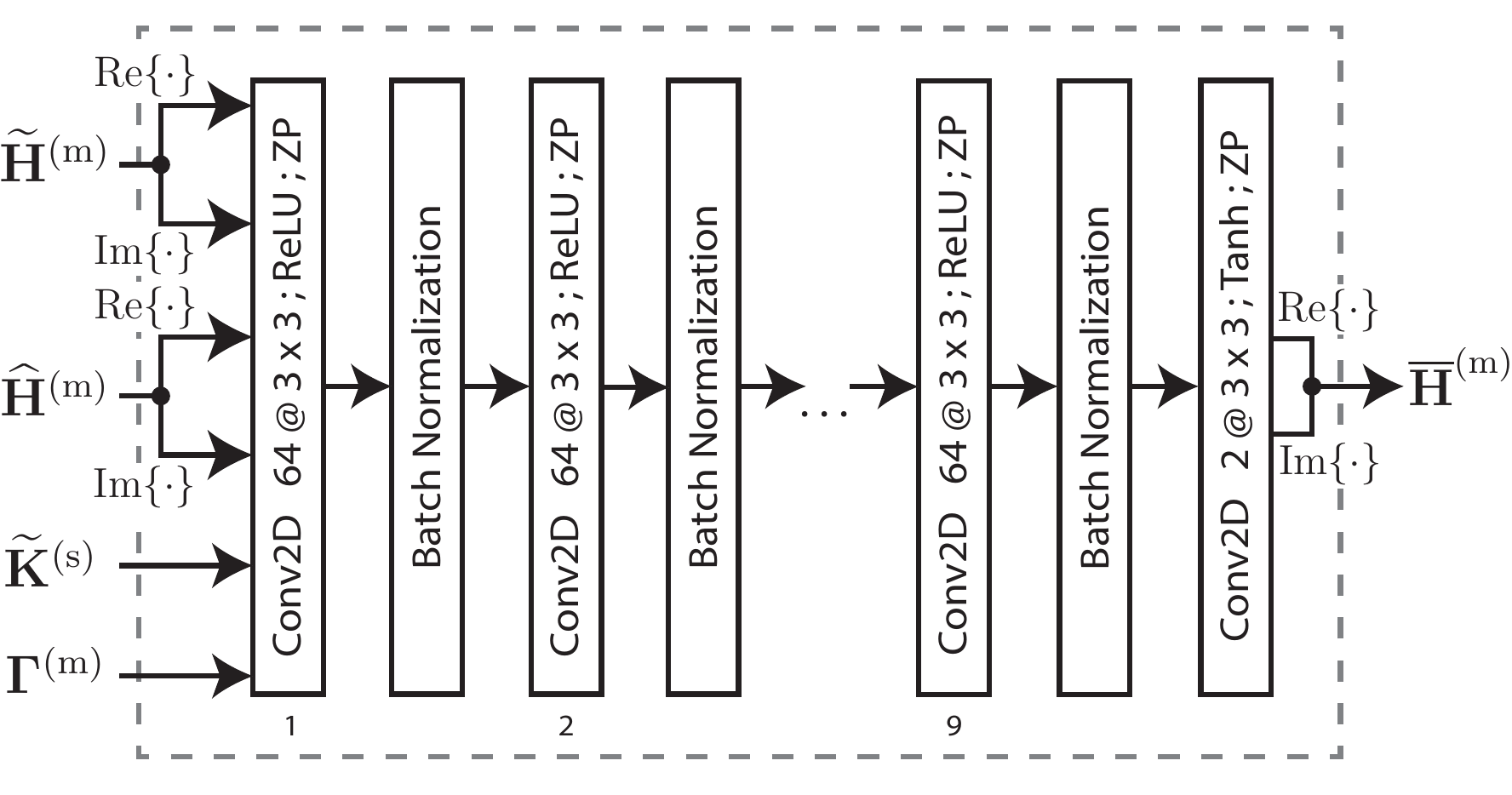}
    \caption{
    The \ac{CNN} architecture comprises nine convolutional layers, followed by an output layer that generates the estimated channel matrix $ \overline{\mathbf{H}}^{\left( {\rm m} \right)} [n]$. 
    }
    \label{fig:block_scheme_cnn}
\end{figure}

\ac{CNN}s are particularly well suited for \ac{MIMO} systems, where the spatial arrangement of antennas plays a crucial role~\cite{Dong2019}.
As input to the \ac{CNN}, we use the in-band $\widetilde{\mathbf{H}}^{\left( \rm m \right)} [n]$ and out-of-band aided $\widehat{\mathbf{H}}^{\left( \rm m \right)} [n]$ channel estimates.
Furthermore, as in~\cref{eq:optimal_w}, we provide as input the estimated sub-6\,GHz $K$-factor matrix $ \widetilde{\mathbf{K}}^{\left( \rm s \right)} = \widetilde{K}^{\left( \rm s \right)} \mathbf{1}_{M_{\rm Rx}^{\left( \rm m \right)}  \times M_{\rm Tx}^{\left( \rm m \right)} }$ and the noise variance matrix $ \mathbf{\Gamma}^{\left( \rm m \right)} = \sigma_{w^{\left( \rm m \right)}}^2 \mathbf{1}_{M_{\rm Rx}^{\left( \rm m \right)}  \times M_{\rm Tx}^{\left( \rm m \right)} }$.
We assume that the noise variance $\sigma_{w^{\left( \rm m \right)}}^2$ is known at the receiver and that is related to pre-beamforming \ac{SNR} with $\gamma^{\left( \rm m \right)} = 1 / \sigma_{w^{\left( \rm m \right)}}^2$.
The \ac{CNN} outputs the estimated channel matrix $ \overline{\mathbf{H}}^{\left( {\rm m} \right)} [n]$ through the mapping relationship
\begin{equation}
    \overline{\mathbf{H}}^{\left( {\rm m} \right)} [n] = f_{\Phi_{\rm C}} \left( 
    \widetilde{\mathbf{H}}^{\left( \rm m \right)} [n], 
    \widehat{\mathbf{H}}^{\left( \rm m \right)} [n], 
    \widetilde{\mathbf{K}}^{\left( \rm s \right)}, 
    \mathbf{\Gamma}^{\left( \rm m \right)} ; \Phi_{\rm C}   \right), 
    \label{eq:cnn_mapping}
\end{equation}
where $\Phi_{\rm C}$ denotes the parameter set of the \ac{CNN}. 
The block diagram of the \ac{CNN} architecture is illustrated in~\cref{fig:block_scheme_cnn}. 
Each intermediate layer employs the \ac{ReLU} activation function, followed by a \ac{BN} layer to stabilize and accelerate the learning process.
Additionally, \ac{ZP} is applied to maintain the dimensions of the feature matrix during convolution. 
The output layer applies a hyperbolic tangent activation function to generate the estimated real and imaginary parts of the channel matrices.

\subsubsection{UNet-based Method} \label{subsubsec:unet_method}
\begin{figure}[t]
    \centering
    \includegraphics[width=\linewidth]{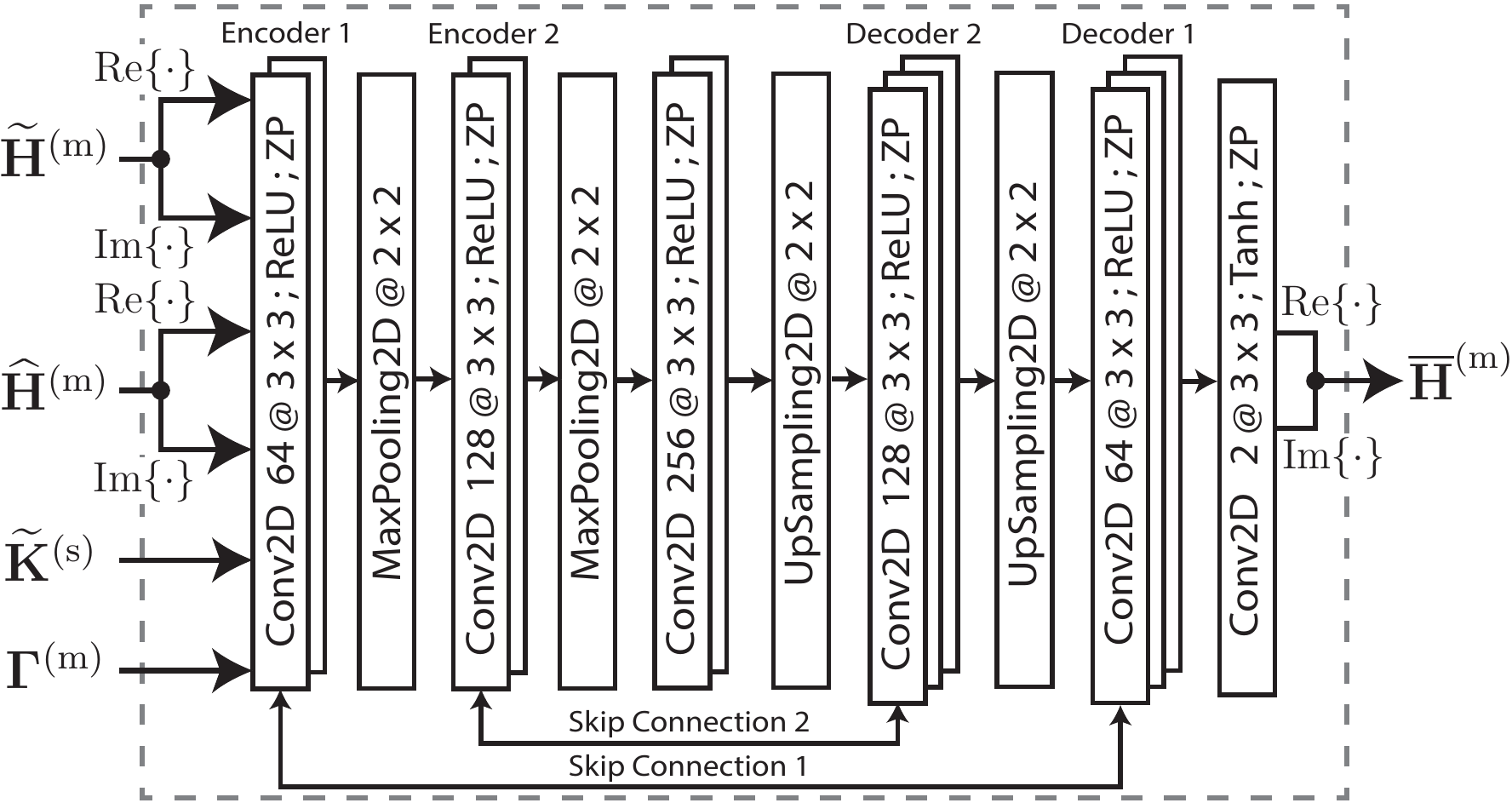}
    \caption{
    The UNet architecture consists of two encoders in the contraction path and two decoders in the symmetric expansion path, followed by an output layer that produces the estimated channel matrix $ \overline{\mathbf{H}}^{\left( {\rm m} \right)} [n]$. 
    }
    \label{fig:block_scheme_unet}
\end{figure}

To further enhance performance in terms of \ac{SE}, we employ UNet-based channel estimation.
UNet offers advantages over standard \ac{CNN}s, due to its ability to capture both global information and fine-grained details via skip connections and its encoder-decoder structure. 
In~\cite{Zhao2023}, UNet demonstrated greater robustness to noise compared to standard \ac{CNN}s.
The UNet takes the same input as the CNN and outputs the estimated channel matrix $ \overline{\mathbf{H}}^{\left( {\rm m} \right)} [n]$ through the mapping relationship
\begin{equation}
    \overline{\mathbf{H}}^{\left( {\rm m} \right)} [n] = f_{\Phi_{\rm U}} \left( 
    \widetilde{\mathbf{H}}^{\left( \rm m \right)} [n], 
    \widehat{\mathbf{H}}^{\left( \rm m \right)} [n], 
    \widetilde{\mathbf{K}}^{\left( \rm s \right)}, 
    \mathbf{\Gamma}^{\left( \rm m \right)} ; \Phi_{\rm U}   \right), 
    \label{eq:unet_mapping}
\end{equation}
where $\Phi_{\rm U}$ denotes the parameter set of the UNet. 
The block diagram of the UNet architecture is shown in~\cref{fig:block_scheme_unet}. 
Each encoder and decoder comprises five convolutional layers, each followed by a \ac{BN} layer.
The contraction path captures contextual information, while the symmetric expansion path enables precise spatial localization~\cite{Zhao2023}.
Skip connections are employed to directly link encoder layers to corresponding decoder layers, enabling the transfer of features extracted during encoding to enhance decoding performance.
At the end of the expansion path, the final decoder block is used to generate the output.
The output layer applies a hyperbolic tangent activation function to generate the estimated real and imaginary parts of the channel matrices.

\subsubsection{Offline Training} \label{subsubsec:cnn_offline_training}
For the proposed networks, the training set consisting of $L_{\rm train}$ samples is generated according to a specific channel model in the simulation environment. 
The $l$-th sample is represented as $     \left( 
        \widetilde{\mathbf{H}}^{\left( \rm m \right)}_l,
        \widehat{\mathbf{H}}^{\left( \rm m \right)}_l, 
        \widetilde{\mathbf{K}}^{\left( \rm s \right)}_l, 
        \mathbf{\Gamma}^{\left( \rm m \right)}_l; 
        \mathbf{H}^{\left( \rm m \right)}_l
        \right),$
where $\widetilde{\mathbf{H}}^{\left( \rm m \right)}_l, \widehat{\mathbf{H}}^{\left( \rm m \right)}_l, \widetilde{\mathbf{K}}^{\left( \rm s \right)}_l$ and  $\mathbf{\Gamma}^{\left( \rm m \right)}_l$ are input data, and $ \mathbf{H}^{\left( \rm m \right)}_l$ is the target data. 
To reduce the complexity of the model, we consider only the single central subcarrier ($n = N^{\rm \left( m \right)} / 2$) for both the input and target data.
Before entering the network, the $\widetilde{\mathbf{H}}^{\left( \rm m \right)}_l$ and $\widehat{\mathbf{H}}^{\left( \rm m \right)}_l$ are separated into their real and imaginary parts.
As input, the \ac{CNN} and UNet take six real-valued matrices of dimension $M_{\rm Rx}^{\rm \left( m \right)} \times M_{\rm Tx}^{\rm \left( m \right)}$.
At the output, the real and imaginary parts are recombined to obtain the $M_{\rm Rx}^{\rm \left( m \right)} \times M_{\rm Tx}^{\rm \left( m \right)}$ complex-valued estimated channel matrix $\overline{\mathbf{H}}^{\left( {\rm m} \right)}_l$.
The goal of the offline training is to minimize the normalized channel estimation \ac{MSE} loss function
\begin{equation} 
    {\rm NMSE_{Loss}} =
    \frac{1}{L_{\rm train} }
    \sum_{l=1}^{L_{\rm train}} 
    \frac{
    \norm{
    \mathbf{H}_l^{\left( {\rm m} \right)}
    - \overline{\mathbf{H}}_l^{ \left( {\rm m} \right)}
    }_{F}^2}
    {\norm{\mathbf{H}_l^{ \left( {\rm m} \right)}}_{F}^2},
    \label{eq:mse_cnn}
\end{equation}
with the networks being optimized using Adam optimization, with a batch size of 64 and a total of 200 training epochs.

\subsubsection{Online Deployment} \label{subsubsec:cnn_online_deployment}
After the offline training, the trained \ac{CNN} and UNet are deployed at the receiver for online channel estimation.
In this phase, the networks trained for a single subcarreir are applied for each subcarrier $n \in \{ 1, \ldots, N^{\left( \rm m \right) }\}$.
As the output, the \ac{CNN} and UNet provide the estimated channel matrix $ \overline{\mathbf{H}}^{\left( {\rm m} \right)} [n]$, generated through the mappings given by~\cref{eq:cnn_mapping} and~\cref{eq:unet_mapping}, respectively.
The networks are robust and trained for different wireless channel conditions, in terms of $K$-factor and \ac{SNR}.
Therefore, there is no need to retrain the networks for each specific $K$-factor or \ac{SNR} scenarios, which saves computing power and storage space.
However, both the \ac{CNN} and UNet are limited with respect to the choice of \ac{MIMO} antenna configuration.
If the network should be applied on a different \ac{MIMO} configuration, retraining with different $M_{\rm Rx}^{\rm \left( m \right)} \times M_{\rm Tx}^{\rm \left( m \right)}$ parameters is necessary.

\section{Simulation-based Comparison} \label{sec:comparison}
To assess the performance of the proposed channel estimation methods, we simulate the normalized channel estimation and interpolation \ac{MSE}, and the achievable \ac{SE} in a frequency-selective channel. 
The simulation parameters are listed in~\cref{tab:simParams}, while the channel is modeled using the \ac{3GPP} urban macro for the \ac{LOS} scenario with an adjustable $K$-factor, as described in~\cite{3gpp.38.901}.
Based on the measurement-validated similarity of $K$-factors reported in~\cite{Pasic2025}, the scaling factor $c_K$ is set to 1, defining the \ac{MMW} $K$-factor as $K^{\left( \rm m \right)}= K^{\left( \rm s \right)}$.

In addition to the \ac{MRC} method discussed in~\cref{subsec:mrc} and the \ac{ML} methods presented in~\cref{subsec:deep_learning}, we show the performance of methods that rely solely on in-band \ac{MMW} information.
Specifically, we include a non-\ac{ML} baseline, where only the \ac{MMW} channel estimate $\overline{\mathbf{H}}^{\left( {\rm m} \right)} [n] = \widetilde{\mathbf{H}}^{\left( {\rm m} \right)} [n] $ is utilized.
Furthermore, we show the performance of in-band \ac{CNN} and in-band UNet methods, which exclude both out-of-band estimate $\widehat{\mathbf{H}}^{\left( \rm m \right)} [n]$ and the $K$-factor $\widetilde{\mathbf{K}}^{\left( \rm s \right)}$.
For these methods, we consider $\overline{\mathbf{H}}^{\left( {\rm m} \right)} [n] = f_{\Phi_{\rm C}} \left( \widetilde{\mathbf{H}}^{\left( \rm m \right)} [n], \mathbf{\Gamma}^{\left( \rm m \right)} ; \Phi_{\rm C} \right)$ for the in-band \ac{CNN} and $\overline{\mathbf{H}}^{\left( {\rm m} \right)} [n] = f_{\Phi_{\rm U}} \left( \widetilde{\mathbf{H}}^{\left( \rm m \right)} [n], \mathbf{\Gamma}^{\left( \rm m \right)} ; \Phi_{\rm U} \right)$ for the in-band UNet.
Additionally, we show the performance for the case of perfect \ac{CSI} $\left( \overline{\mathbf{H}}^{\left( {\rm m} \right)} [n] = \mathbf{H}^{\left( {\rm m} \right)} [n] \right)$ to establish an upper performance bound.

\begin{table}[t]
    \centering
    \caption{Simulation Parameters} 
    \label{tab:simParams}
    \begin{tabular}{rcc}
        \hline
        \textbf{Parameter}                          & \multicolumn{2}{c}{\textbf{Value}} \\ \hline
        Frequency Band                              & sub-6 GHz         & mmWave         \\
        Carrier Frequency $f_{\rm c}$               & 2.55\,GHz         & 25.5\,GHz           \\
        Wavelength $\lambda$                        & 11.76\,cm         & 1.176\,cm          \\
        MIMO Configuration                          & 8$\times$8        & 8$\times$8        \\
        Bandwidth $B$                               & 20.16\,MHz        & 403.2\,MHz            \\
        Subcarrier Spacing $\bigtriangleup f$       & 60\,kHz           & 120\,kHz             \\
        Cyclic Prefix $t_{\rm CP}$                  & 1.19\,$\mu$s      & 0.59\,$\mu$s             \\ \hline    
    \end{tabular}
\end{table}

The normalized channel estimation and interpolation \ac{MSE} is represented by
\begin{equation} 
    \mathrm{NMSE} = \frac{1}{L_r N^{\left( \rm m \right)}}
    \sum_{l=1}^{L_r} \sum_{n=1}^{N^{\left( \rm m \right)}} 
    \frac{
    \norm{
    \mathbf{H}_l^{\left( {\rm m} \right)} [n]
    - \overline{\mathbf{H}}_l^{ \left( {\rm m} \right)} [n]
    }_{F}^2}
    {\norm{\mathbf{H}_l^{ \left( {\rm m} \right)} [n]}_{F}^2},
\label{eq:mse}
\end{equation}
where $L_r$ denotes the number of realizations and $l$ indicates a specific channel~realization.
The achievable \ac{SE}, averaged over $N^{\left( \rm m \right)}$ subcarriers and expressed in bits$/$s$/$Hz, is defined by
\begin{equation} 
    \mathrm{SE}= 
    \frac{1}{N^{\left( \rm m \right)}} 
    \sum_{n=1}^{N^{\left( \rm m \right)}} 
    \sum\limits_{\mu=1}^{\ell_{\rm max}} 
    \log_2 \left( 1 + \mathrm{SINR}_{\rm \mu} [n] \right),
    \label{eq:se}
\end{equation}
where the effective \ac{SINR} for the stream ${\rm \mu}$ is given by~\cite{Pasic2024_TCOM}
\begin{equation} 
    \mathrm{SINR}_{\rm \mu} [n] = \frac{ 
    \left| \overline{\mathrm{G}}_{\rm \mu,\mu}^{\left( {\rm m} \right)} [n] \right|^2 }
    { \sum\limits_{\substack{\nu=1 \\ \nu \neq \mu}}^{\ell_{\rm max}} 
    \left| \overline{\mathrm{G}}_{\rm \mu, \nu}^{\left( {\rm m} \right)} [n] \right|^2
    + \left( \sigma_{\rm \mu}^2 [n] + \sigma_{w^{\left( \rm m \right)}}^2 \right) 
    \norm{ \overline{\mathrm{Q}}_{\rm :,\mu}^{\left( {\rm m} \right)} [n] }^2 }.    
    \label{eq:sinr}
\end{equation}
In~\cref{eq:sinr}, $\overline{\mathrm{G}}_{\rm \mu,\nu}^{\left( {\rm m} \right)} [n]$ with $\mu, \nu \in \{ 1, \ldots, \ell_{\rm max} \} $ represents the elements of the channel gain matrix $\overline{\mathbf{G}}^{\left( {\rm m} \right)} [n] \in \mathbb{C}^{\ell_{\rm max} \times \ell_{\rm max}}$ for the $n$-th subcarrier, defined as
\begin{equation}
    \overline{\mathbf{G}}^{\left( {\rm m} \right)} [n]= 
    \left( \overline{\mathbf{Q}}^{\left( {\rm m} \right)} [n] \right) ^ {\rm H} 
    \mathbf{H}^{\left( {\rm m} \right)} [n]
    \overline{\mathbf{F}}^{\left( {\rm m} \right)} [n] 
    \left( \overline{\mathbf{P}}^{\left( \rm m \right)} [n] \right)^{1/2}.
    \label{eq:chgain}
\end{equation}
Moreover, $ \sigma_{\rm \mu}^2 [n]$ denotes the diagonal entries (variance) of the estimation error covariance 
$    \overline{\mathbf{C}}_{\varepsilon}^{\left( {\rm m} \right)} [n] = 
    \frac{1}{\ell_{\rm max}}
    \overline{\boldsymbol{\varepsilon}}^{\left( {\rm m} \right)} [n] 
    \left( \overline{\boldsymbol{\varepsilon}}^{\left( {\rm m} \right)} [n] \right)^{\rm H}
$
for the $n$-th subcarrier.
The error $\overline{\boldsymbol{\varepsilon}}^{\left( {\rm m} \right)} [n]$ quantifies the difference between the channel gain matrix computed with the estimated precoder/combiner
$\overline{\mathbf{G}}^{\left( {\rm m} \right)} [n]$, and the channel gain matrix obtained using the ideal precoder/combiner $\mathbf{G}^{\left( {\rm m} \right)} [n]$.

\begin{figure}[t]
    \centering
    {\includegraphics[width=\columnwidth]{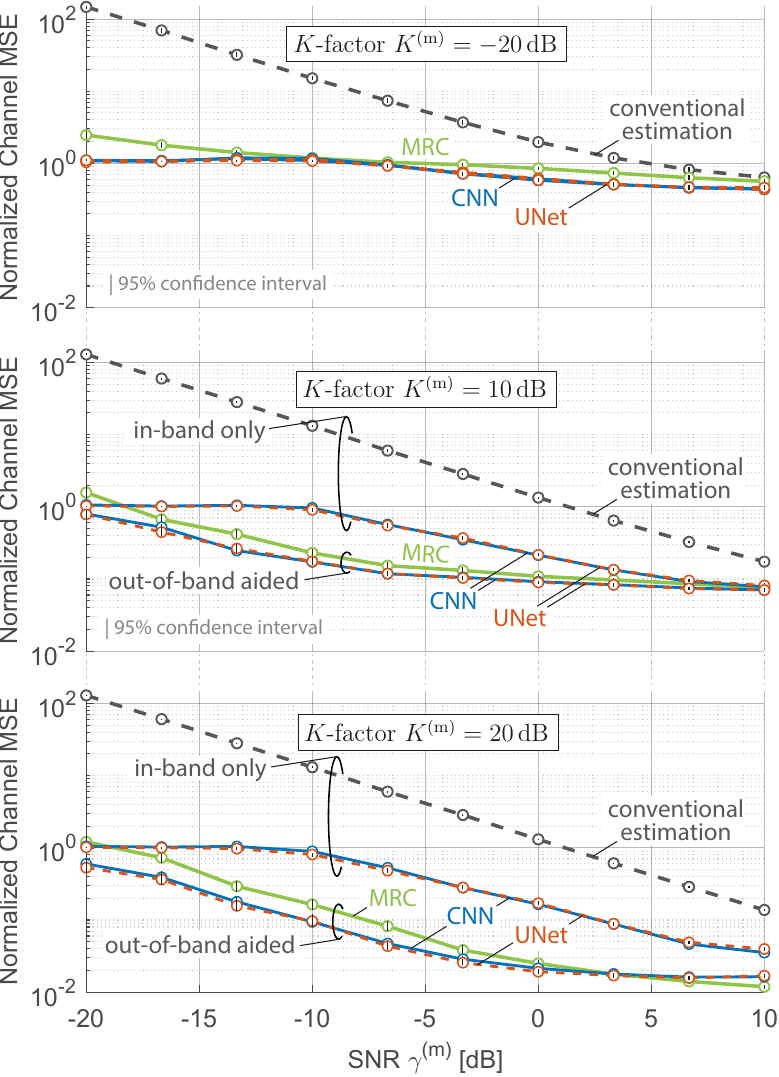}}
    \caption{
    The proposed out-of-band aided CNN and UNet channel estimation methods achieve the lowest channel \ac{MSE} across all $K$-factors. 
    The small vertical bars within the circular markers indicate the 95\% confidence intervals.
    }
   \label{fig:mse}
\end{figure}

In the first simulation, we analyze performance as a function of \ac{SNR}, with results shown in~\cref{fig:mse} for $L_r=100$ realizations in terms of normalized channel \ac{MSE}.
The normalized channel \ac{MSE} decreases with increasing \ac{SNR}, regardless of the estimation method or $K$-factor.
Moreover, all estimation methods consistently outperform the conventional non-\ac{ML} approach across the entire \ac{SNR} range and for all $K$-factors.

\textbf{Results at K-factor of $-$20\,dB:}
In this scenario, the in-band and out-of-band aided \ac{ML} methods exhibit comparable performance, all outperforming the \ac{MRC} method across the entire \ac{SNR} range.

\textbf{Results at K-factor of 10\,dB:}
The \ac{MRC} method outperforms the in-band \ac{CNN} and UNet methods, highlighting the significance of missing out-of-band information.
However, out-of-band aided \ac{CNN} and UNet methods maintain comparable performance, outperforming the \ac{MRC} at all \ac{SNR} levels.

\textbf{Results at K-factor of 20\,dB:}
While the \ac{MRC} method outperforms the in-band \ac{CNN} and UNet methods, it underperforms relative to the out-of-band-aided \ac{CNN} and UNet approaches. 
The out-of-band-aided \ac{ML} methods demonstrate comparable performance across all \ac{SNR} levels.

\begin{table}[t]
    \centering
    \caption{Dataset Parameters} 
    \label{tab:datasetParams}
    \begin{tabular}{rcc}
        \hline
        \textbf{Parameter}                          & Value \\ \hline
        Training Samples $L_{\rm train}$,             & 500\,000         \\
        Test Samples $L_{\rm test}$                 & 500\,000         \\
        SNR $\gamma^{\left( \rm m \right)}$ [dB]    & $\mathcal{U} \sim$ [$-$20, 10] \\
        $K$-factor $K^{\left( \rm m \right)}$ [dB]  & $\mathcal{U} \sim$ [$-$20, 30] \\
        AoD $\vartheta$, AoA $\varphi$ [deg]        & $\mathcal{U} \sim$ [$-$90, 90] \\ \hline
    \end{tabular}
\end{table}

\begin{figure}[t]
    \centering
    {\includegraphics[width=\columnwidth]{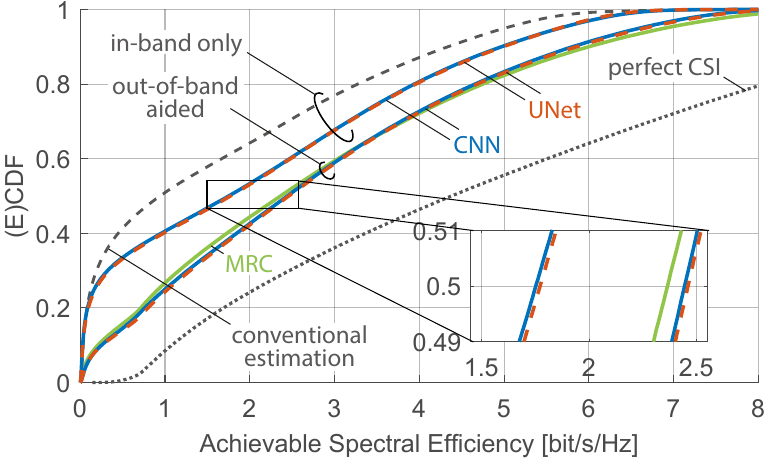}}
    \caption{
    The proposed out-of-band aided CNN and UNet channel estimation methods achieve a 3.35\% and 4.03\% higher median \ac{SE}, respectively, compared to the \ac{MRC} method.
    }
   \label{fig:se_cdf}
\end{figure}

In the next simulation, we generate a dataset with $L_{\rm train}$ and $L_{\rm test}$ channel realizations.
For each realization, the \ac{SNR} $\gamma^{\left( \rm m \right)}$, $K$-factor $K^{\left( \rm m \right)}$, \ac{AoD} $\vartheta$ and \ac{AoA} $\varphi$ are mutually independent and uniformly distributed within the ranges specified in~\cref{tab:datasetParams}.
The \ac{CDF} of the achievable \ac{SE} for all $L_{\rm test}$ test samples is shown in~\cref{fig:se_cdf}.
For in-band methods, the \ac{CNN} and UNet architectures achieve 86\% and 88\% higher median \ac{SE}, respectively, than the non-\ac{ML} conventional method.
However, substantial gains in achievable \ac{SE} are observed with out-of-band aided channel estimation.
The \ac{MRC} achieves 151\% higher median \ac{SE}, increasing to 159\% with the out-of-band aided \ac{CNN} and 161\% with the out-of-band aided UNet.
Overall, in both in-band only and out-of-band aided cases, the UNet slightly outperforms the \ac{CNN} architectures in terms of median \ac{SE}.
However, this gain in \ac{SE} comes at the cost of increased computational complexity due to the more complex design of the UNet.

\section{Conclusion} \label{sec:conclusion}
In this paper, we propose two novel deep-learning channel estimation methods for \ac{MMW} \ac{MIMO} systems that leverage sub-6\,GHz out-of-band information.
The proposed \ac{CNN} and UNet architectures, when combined with out-of-band information, increase median achievable \ac{SE} by approximately 3-4\% compared to the non-\ac{ML} \ac{MRC} method.
While both architectures exhibit comparable performance on a large scale, the UNet architecture demonstrates a slight advantage  over the \ac{CNN}, due to its ability to capture fine-grained details.

\bibliography{references}
\bibliographystyle{IEEEtran}

\end{document}

%% file: acronyms.tex
\begin{acronym}
\acro{AWG}[AWG]{Arbitrary Waveform Generator}
\acro{CW}[CW]{Continuous Wave}
\acro{HST}[HST]{High-Speed Train}
\acro{IF}[IF]{Intermediate Frequency}
\acro{LO}[LO]{Local Oscillator}
\acro{LSF}[LSF]{local scattering function}
\acro{MMW}[mmWave]{millimeter wave}
\acro{OFDM}[OFDM]{orthogonal frequency-division multiplexing}
\acro{PCB}[PCB]{Printed Circuit Board}
\acro{SMD}[SMD]{Surface Mount Device}
\acro{SNR}[SNR]{signal-to-noise ratio}
\acro{SINR}[SINR]{signal-to-interference-and-noise ratio}
\acro{RF}[RF]{radio frequency}
\acro{V2X}[V2X]{vehicle-to-everything}
\acro{IC}[IC]{Integrated Circuit}
\acro{FPGA}[FPGA]{Field Programmable Gate Array}
\acro{ISI}[ISI]{Inter-Symbol Interference}
\acro{CSIT}{channel state information at the transmitter}
\acro{CSI}{channel state information}
\acro{ML}{machine learning}
\acro{RNN}{recurrent neural networks}
\acro{LSTM}{long short-term memory}
\acro{GRU}{Gated Recurrent Unit}
\acro{FDD}{frequency-division duplex}
\acro{TDD}{time-division duplex}
\acro{CIR}{channel impulse response}
\acro{CTF}{channel transfer function}
\acro{PDP}{power delay profile}
\acro{DSD}{Doppler power spectral density}
\acro{IFFT}{Inverse Fast Fourier Transform}
\acro{ITS}[ITS]{intelligent transportation systems}
\acro{5G}[5G]{fifth generation}
\acro{NR}[NR]{new radio}
\acro{QAM}[QAM]{quadrature amplitude modulation}
\acro{ICI}[ICI]{Inter-Carrier Interference}
\acro{MSE}[MSE]{mean squared error}
\acro{BER}[BER]{bit error ratio}
\acro{RMS}[RMS]{root-mean-square}
\acro{TDL-A}[TDL-A]{tapped delay line A}
\acro{TDL-D}[TDL-D]{tapped delay line D}
\acro{ISI}[ISI]{Inter-Symbol Interference}
\acro{DFT}[DFT]{discrete Fourier transform}
\acro{IDFT}[IDFT]{inverse discrete Fourier transform}
\acro{CTF}[CTF]{channel transfer function}
\acro{DR}[DR]{dynamic range}
\acro{HPBW}[HPBW]{half-power beamwidth}
\acro{DPSS}[DPSS]{discrete prolate spheroidal sequences}
\acro{CDF}[CDF]{cumulative distribution function}
\acro{URLLC}[URLLC]{ultra-reliable low-latency communication}
\acro{3GPP}[3GPP]{3rd Generation Partnership Project}
\acro{MIMO}[MIMO]{multiple-input multiple-output}
\acro{SISO}[SISO]{single-input single-output}
\acro{MRC}[MRC]{maximal ratio combining}
\acro{AWGN}[AWGN]{additive white Gaussian noise}
\acro{LS}[LS]{least-squares}
\acro{SVD}[SVD]{singular value decomposition}
\acro{SE}[SE]{spectral efficiency}
\acro{ULA}[ULA]{uniform linear array}
\acro{FSPL}[FSPL]{free space path loss}
\acro{LOS}[LOS]{line-of-sight}
\acro{NLOS}[NLOS]{non-line-of-sight}
\acro{AoA}[AoA]{angle-of-arrival}
\acro{MUSIC}[MUSIC]{multiple signal classification}
\acro{EVD}[EVD]{eigenvalue decomposition}
\acro{AoD}[AoD]{angle-of-departure}
\acro{RMSE}[RMSE]{root mean squared error}
\acro{SIMO}[SIMO]{single-input multiple-output}
\acro{MAC}[MAC]{medium access control}
\acro{eMBB}[eMBB]{enhanced mobile broadband}
\acro{SVM}[SVM]{support vector machine}
\acro{VNA}[VNA]{vector network analyzer}
\acro{CNN}[CNN]{convolutional neural network}
\acro{DNN}[DNN]{deep neural network}
\acro{ReLU}[ReLU]{rectified linear unit}
\acro{ZP}[ZP]{zero padding}
\acro{BN}[BN]{batch normalization}
\acro{NN}[NN]{neural network}
\acro{ADAM}[ADAM]{adaptive moment estimation}
\end{acronym}